\documentclass[conference]{IEEEtran}

\usepackage[utf8]{inputenc}
\usepackage[T1]{fontenc}

\usepackage{hyperref}
\usepackage{footnote}

\usepackage[
backend=bibtex, 
style=ieee, 
firstinits=true,
doi=false,
isbn=false,
url=false
]
{biblatex}

\iflanguage{ngerman}{
\newcommand{\q}[1]{„#1“}

}{ 
\newcommand{\q}[1]{``#1''}

}

\usepackage{booktabs}
\usepackage{xcolor}
\usepackage{framed}

\newlength{\leftbarwidth}
\newlength{\leftbarsep}
\colorlet{leftbarcolor}{black}

\renewenvironment{leftbar}{%
    \MakeFramed {\advance \hsize -\width \FrameRestore }%
}{%
    \endMakeFramed
}

\usepackage{tabularx}

\usepackage{amssymb}

\bibliography{textanalysis}

\usepackage{amsmath,amssymb,amsfonts}
\usepackage{algorithmic}
\usepackage{graphicx}
\usepackage{textcomp}
\def\BibTeX{{\rm B\kern-.05em{\sc i\kern-.025em b}\kern-.08em
    T\kern-.1667em\lower.7ex\hbox{E}\kern-.125emX}}
\begin{document}

\title{\huge Text Classification Components for Detecting\\Descriptions and Names of CAD models}

\author{\IEEEauthorblockN{Thomas Köllmer}
\IEEEauthorblockA{\textit{Fraunhofer Institute for}\\\textit{Digital Media Technology IDMT}\\
D-98693 Ilmenau\\
thomas.koellmer@idmt.fhg.de}
\and
\IEEEauthorblockN{Jens Hasselbach}
\IEEEauthorblockA{\textit{Fraunhofer Institute for}\\\textit{Digital Media Technology IDMT}\\
D-98693 Ilmenau\\
jens.hasselbach@idmt.fhg.de}
\and
\IEEEauthorblockN{Patrick Aichroth}
\IEEEauthorblockA{\textit{Fraunhofer Institute for}\\\textit{Digital Media Technology IDMT}\\
D-98693 Ilmenau\\
patrick.aichroth@idmt.fhg.de}}
\maketitle

\begin{abstract}

We apply text analysis approaches for a specialized search engine for 3D CAD models and associated products.
The main goals are to distinguish between actual product descriptions and other text on a website, as well as to decide whether a given text is or 
contains a product name.

For this we use paragraph vectors for text classification, a character-level long short-term memory network (LSTM) for a single word classification and an LSTM tagger based on word embeddings for detecting product names within sentences.
Despite the need to collect bigger datasets in our specific problem domain, the first results are promising and partially fit for production use.

\end{abstract}


\section{Introduction}

Text analysis is a crucial processing step when collecting relevant and filtering out non-relevant content from websites. 
The context of this paper is the prototype of a search engine specialized for the retrieval of 3D CAD models (computer-aided design).
Manufacturers of furniture, e.g., for offices, provide CAD files on their websites to be used by architects in their planning tools.
However, often those files are hard to find on their web pages, e.g., hidden in a separate download area.
What makes matters even worse is that every manufacturer's website has its own structure. 
This makes it hard to quickly find the required models and invites the user to download a model only once and miss updates in the progress.

This work is part of an umbrella project with the goal to not only make various manufacturers' sites searchable (this can be done by all major search engines and the manufacturers sites themselves), but also to present products in a unified way, joining product description texts, pictures and also CAD-files in one coherent interface. 
The diversity of different manufacturers website presentations and the fact that product descriptions are mostly separate from their associated CAD files, makes this challenging and requires a good understanding how the products are organized within a website.

Given the amount of manufacturers and models, it is not feasible to hand-craft crawlers for every manufacturer, nor is there a widely accepted standard to indicate the information we need in a machine-readable format.
Using a combination of heuristics and automatic content analysis (text analysis, image analysis, CAD model analysis) we need to find a way to extract the information at an acceptable quality level with minimal human intervention.

This paper puts a spotlight on textual analysis problems that arise in such an endeavour.
First, we detect whether a piece of text resembles a product description. 
Second, we decide if a given text is or contains a product name.
Both are important for summarizing product information and linking product pages with their associated CAD model files, e.g., by searching the product name in a zip archive consisting of CAD files.

While the overall system honours the layout of the page, hints given by the markup or the URL, the goal of the approaches discussed here is to operate on plain text only. 
That way, they can be used as supporting components for the heuristics that analyse the layout of the page. 
Also, they are still useful in cases where there is no clear distinction layout-wise at all.
Combined with optical features (color, capitalization, placement on the page) the proposed techniques help finding the right product name, but also detecting false positives, that might be suppressed from the result view, or can  be flagged for manual classification.

In the scope of this paper, we do not apply dictionary approaches, e.g., filtering imprints and company names, but try to build general purpose classifiers using machine learning. 

\section{Related Work}
\label{sec:related_work}

A key innovation for text analysis was the approach of using unsupervised learning to create word embeddings, that capture the semantics of words surprisingly well~\cite{DBLP:journals/corr/abs-1301-3781}.

Text classification traditionally depends on calculating statistics on input text, e.g., using Naive Bayes approaches or Support Vector Machines (SVM). But also in this domain, neural networks are competitive, either word based~\cite{lai2015recurrent} or character based~\cite{zhang2015character}.

Part of speech (POS) tagging made a huge leap forward in the last years using recurrent neural networks (RNN) instead of hand crafted features, or combining both.
The winning approach of the 2017 ConLL task on part of speech tagging is based on LSTMs (Long short-term memory networks, a recurrent neural network architecture)~\cite{dozat-qi-manning:2017:K17-3}.
In fact, all but one approaches from the top 10 of this competition are based on recurrent neural networks, most of the time a bidirectional LSTM, either based on word embeddings or character level representations \cite{santos2014learning}.

These approaches have in common that they perform excellent given enough training data. 
For our current research, this is not directly applicable as we do not have annotated data samples of product description texts in that order of magnitude for all required languages. 
We will try a best effort approach to use the practically proven application of LSTMs to text analysis for our problem domain.


\section{Data Sets}
\label{sec:data_sets}

\begin{table}[t]
\centering

\caption{Statistics for selected collections. Descriptions must be longer than 200 character and contain a title with more than 6 characters to be included in the filtered data.}

\begin{tabularx}{\columnwidth}{Xrrr}

\toprule
Category     &  Overall   & Filtered   &   Ratio    \\
\midrule
Office Products     &   134,838   &  1,595         &  1.18\%            \\
Movies and TV       &   208,321   &  5,150         & 2.47\%             \\
CDs and Vinyl       &   566,931   &  16,703       & 2.95\%              \\
Tools and Home Improvement   &  269,120    & 1,791   &  0.67\%  \\
Grocery and Gourmet Food   &    171,760    &  3,483  & 2,03\%  \\
Sports and Outdoors    &        532,197    &  13,039   &  2,45\%  \\

\bottomrule                   
\end{tabularx}

\label{tab:categories}

\end{table}

As discussed in the previous section, given huge amounts of categorized text, say from Wikipedia or one of the larger treebanks, e.g., from the Universal Dependencies project, 
the task of general text classification is more or less solved for many practical purposes.
However, for our exact problem, there are not so many suitable datasets available in different languages.

For the task of finding product names in product descriptions, we use the Amazon Product dataset by McAuley~\cite{He:2016:UDM:2872427.2883037}, which focusses on product reviews, but also contains product descriptions and product names from various categories like Books or Office Products.
For example, the Office Product category, which might come close to our problem domain contains 134,838 products. After excluding samples that do not contain a variation of the product name in the product description or fall below a certain length this amount decreases drastically, but still gives a good starting point for training and testing our classifiers.  Table~\ref{tab:categories} shows the number of relevant descriptions for some selected categories.

Besides general purpose datasets for a baseline testing of the classifiers, e.g., the 20Newsgroups dataset\footnote{Available at \url{http://qwone.com/~jason/20Newsgroups/}}, we chose to create our own dataset: Text snippets directly retrieved by the crawling component were hand-tagged being a product description, something else or undecided. This dataset includes 2,920 samples of relatively short text (10 to 30 words) from seven different manufacturers, all crawled by our search engine and therefore directly from the problem domain.

Table~\ref{tab:examplelabel} shows three representatives examples of snippets that needs classification. 
Note, that not in all cases it is possible to unambiguously decide, whether a given text is a product description or a part of it. 
While the first example is most certainly part of a description, and the second example not -- for the third it is not easy to tell, as it totally depends on the context the snippet was retrieved from. 
For training the classifier, we only use the first two categories.

\begin{table}[t]

\caption{Example classifications for the product description detection. Not all classifications are unambiguous.}

\begin{tabularx}{\columnwidth}{>{\textsf}X>{\bfseries}l}
\toprule
Homely atmosphere, sleek elegance for individual and team workstations, solid wood table legs, split sliding top & description\\
\midrule
Creativity works. This was the slogan at this year´s Orgatec office furniture show in Cologne.  &  other\\
\midrule
Our Plenar2 flex cantilever chair also impressed the German Design Award 2015 jury with its outstanding comfort.  &  undecided\\
\bottomrule
\end{tabularx}

\label{tab:examplelabel}
\end{table}

\section{Product Description Detection} 
\label{sec:product_description_detection}

Given the dataset previously discussed, we need a classifier that is able to reliably detect product descriptions. 
We are not overly concerned with the ambiguous cases.
From a system perspective it is more important to detect non-description texts rather than deciding whether the third example needs to be a part of a product description or not.

As this seemed to be a fairly standard text-classification task (e.g., thinking of discriminating texts between two newsgroups as a similar task), we tried established text classification approaches on our data. 
With increasing complexity and novelty, we tried a Naive Bayes classifier, an SVM and two approaches based on Paragraph Vectors, an enhancement of word vector embeddings to capture whole texts, proposed by Le and Mikolov~\cite{le2014distributed}. 
The two latter approaches were implemented on top of deeplearning4j\footnote{\url{https://deeplearning4j.org}}, the other two are based on jLibSvm and the Java-Naive-Bayes-Classifier package.

Table~\ref{tab:evalresults} shows the average result of a 5-fold cross-validation over all samples.
Given that some of the classified samples are very brief and a distinction between two classes is not always perfectly clear, the F1 score of 89\% for the Paragraph Vectors approach is already satisfying for practical applications. The discrimination into more than two classes is not required for our application right now.

Another application of the description detection is to find split points in longer composite text. 
If for some reason the product description and the website footer get concatenated by the crawler, using a sliding window technique can detect such joined texts.

\begin{table}[t]
\centering

\caption{Evaluation results of the text classification on the product description dataset}

\begin{tabularx}{\columnwidth}{Xrrrr}
\toprule
Model                                         &    Precision      &       Recall            &    Accuracy         & F1                \\ \midrule
Naive Bayes\textsuperscript{1}                &    \textbf{0.93}  &       0.72              &    0.83             &  0.81             \\
SVM\textsuperscript{2}                        &    0.74           &       0.62              &    0.70             &  0.67             \\
Paragraph Vector\textsuperscript{3}           &    0.85           &       \textbf{0.94}     &    \textbf{0.88}    & \textbf{0.89}     \\
Recurrent Neural Network\textsuperscript{3}   &    0.85           &       0.89              &    0.87             & 0.87              \\ \bottomrule
\addlinespace[0.2em]
\multicolumn{5}{l}{\textsuperscript{1} \footnotesize{Java-Naive-Bayes-Classifier (1.0.5)}} \\
\multicolumn{5}{l}{\textsuperscript{2} \footnotesize{jLibSVM (based on libsvm 2.88)}} \\
\multicolumn{5}{l}{\textsuperscript{3} \footnotesize{DL4J (1.0.0-alpha)}} \\
\end{tabularx}%

\label{tab:evalresults}
\end{table}






\section{Product Name Detection}
\label{sec:product_name_detection}

Detecting whether a given text is a product description is one thing, a more challenging task is to decide if a given sentence contains a product name, or to decide how likely it is that a single word or group of words resemble a product name.

Let's look at a product description from a well-known furniture store from Sweden for a product named \emph{byholma}: 
\begin{leftbar}
\noindent
\q{The \underline{\smash{byholma}} armchair is handmade from natural fibres and therefore unique, with rounded shapes and nicely detailed patterns.}
\end{leftbar}

\noindent
Intuitively, there are three cues to identify the product name:
\begin{enumerate}
  \item it is followed by the word \q{armchair},
  \item it is part of the subject of a product description text,
  \item the word follows a certain branding aspect, in the concrete case it is a foreign (sounding) word.
\end{enumerate}

As mentioned in the introduction, we will only focus on the raw text features and discard any information on text style and capitalization, information that can be incorporated at another step in the processing chain\footnote{In the spirit of end-to-end deep learning, it might be worthwhile to skip this preprocessing step and learn with all features available. But for now we want a baseline evaluation for cases where there might be no additional features.}.

The first two should be solvable using state-of-the-art methods of part of speech tagging, the third might be an application for character based recurrent neural networks.
In the concrete example: while Byholma is a Swedish village, this information does not matter -- as the branding aspect is being Swedish sounding. 
This is not restricted to natural language. 
Product numbers might be in a consistent format, or fantasy names share phonetic features: \emph{Siamo},  \emph{Amico},  \emph{Sento} and \emph{Previo} are chairs manufactured by the same company.


In the next section we explore how recurrent neural networks can be used to classify words based on their characters, without any information derived from a dictionary or surrounding text. Afterwards we try to identify product names based on their surrounding text.


\subsection{Single Word Classification}






The goal of this classification task is to learn a character-level representation of product names of a certain company. 
Considering the background of this project, we do not consider this as an $n$-class-classification problem, $n$ being the number of manufactures, as it won't be a common case to distinguish names from different companies unless they have joint websites which is uncommon.
Instead we will train a 2-class classifier for every manufacturer. 
It has to decide whether a given word matches the learned representation, i.e., whether it is likely to be a product or not.

A baseline can be made using a dataset from the Swedish furniture company. 
The task is to discriminate their product names from random German family names\footnote{Taken from the PyTorch documentation \url{https://download.pytorch.org/tutorial/data.zip}}.
Both categories are similar in text length, and being both Germanic languages, not too easy to distinguish, given that the classifier is not dictionary based.

To assess the difficulty of that task, a short survey of the authors was made. We asked six colleagues to classify 100 names from our test set: they arrive at an accuracy of around $80\%$ to $90\%$ -- if they are German native speaker. For a non native speaker, the task was much harder. 
This leads to the assumption that they are good at classifying whether a name comes from their language, but bad at classifying the product names.

A vanilla LSTM network achieves an average precision of $84\%$ at cross validation (see table~\ref{tab:ikea}). 
For the selected hyper parameters, see table~\ref{tab:hyper} on the next page.
That the LSTM performed roughly at the same level is somewhat surprising, given that it had no dictionary-like information like humans and does not have a \q{native-speaker-advantage} or a familiarity with a certain brand.

This approach certainly won't work for all manufacturers, but we are optimistic that if there is some branding information put into how the words sound, the word based classification will be able to detect it.
Viewed from another angle: if a list of products is available for a specific manufacturer, the classifier can be trained on it, to predict new products more reliably. Given such lists, it is also easy to evaluate whether this approach works, or if the naming scheme is too much inconsistent.

Given the results from this experiment and the success of character based RNNs in other works, we are optimistic that this approach is well suitable to use in our product search.

\begin{table}[h]
\centering

\caption{Comparison of character based product name detection and human performance.}
\label{tab:ikea}

\begin{tabularx}{\columnwidth}{Xrrrr}

\toprule
Category            &  Precision    & Recall   &   Accuracy    & F1    \\
\midrule
LSTM-Char     &   0.84           &    0.86         &      0.85       &    0.85  \\
German native speaker &  0.85 & 0.87 &   0.84 &   0.85 \\

\bottomrule                                
\end{tabularx}                   
\end{table}

\subsection{Part of Speech Tagging}
\label{sec:pos}

Instead of looking at words separately, handling case 3) from the product name cues, it will be worthwhile to have a look at whole sentences to address the other two hints: the position in the sentence and the presence of indicator words, such as \q{armchair}.

Similar to the single word classification, recurrent neural networks seem to be a good fit, too, as they are able to capture peculiarities of word surroundings. Compared to the general purpose systems on treebank data, we fall short in terms of training data by magnitudes. To mitigate this we chose to evaluate on the Amazon dataset outlined in table~\ref{tab:categories} and optimized our network on the \q{Office Products} category.

Regarding the ground truth quality, this dataset is far from perfect, but gives a first indicator whether a product name detection using techniques from part of speech tagging is a valid approach.

In a first study, we used pre-trained word vectors from the GloVE Dataset~\cite{pennington2014glove} as an input to an LSTM with configuration depicted in table~\ref{tab:hyper}. 
The network labels products on a per-word-basis, which is used to calculate precision, recall and accuracy. 
Note that the use of pre-trained word vectors implies that fantasy words or words borrowed from another language (say, Swedish) will not be in the dataset and get a special vector, all-zeros in our case. 
This is similar to filtering words that are not in the input for the training of the word vectors, which is beneficial as unknown words might indicate a product name.
The results are shown in table~\ref{tab:pos-results}.







\begin{table}[t]
\centering

\caption{Classification results for LSTM based classification based on GloVE data. 5-fold cross validation, Neural network is optimized for Office Products.}
\label{tab:pos-results}

\begin{tabularx}{\columnwidth}{Xrrrr}

\toprule
Category            &  Precision    & Recall   &   Accuracy    & F1    \\
\midrule
Office Products     &   0.34           &    0.83         &      0.92       &    0.48  \\
\midrule
Movies and TV       &   0.09           &  0.54           &       0.90      &    0.16  \\
CDs and Vinyl       &   0.09           &  0.49           &     0.91        &    0.15   \\
Tools and Home Improvement   &   0.32    &    0.91         &     0.91     &   0.47    \\
Grocery and Gourmet Food     &   0.46    &     0.90      &   0.91          &  0.60      \\
Sports and Outdoors      &     0.38     &    0.88      &     0.94        &    0.53     \\

\bottomrule                                
\end{tabularx}                   
\end{table}

For office products, on average 4\% of the words in a description are product names, which we accounted for in the loss function. 
This has the implication that an all-zero classification reaches an accuracy of 0.96; an all-one (i.e., all product) classification 0.04 accuracy and an F1 measure of 0.077. 
While the results leave room for improvement, the classifier is clearly not guessing.
The product categories in the dataset are not homogeneous, the hyper parameters that work well for one category do not work well in another, like CDs and Vinyl, but surprisingly better in other categories (Groceries, Sports).

Considering these results, we decided to focus on getting test data that is more suited to our our main use case  and expand our training set for product descriptions.
Second, we will try a transfer learning approach that starts with a working POS-tagger that is able to capture sentence structures and apply our restriction on tagging product names on top of that.
Third, we will have to combine the results from the word-based classifier with the sentence based classifier.

\section{Conclusion and Outlook}
\label{sec:conclusion_and_outlook}

In this work, we presented text analysis tasks needed to successfully implement a search engine for 3D CAD models from furniture manufacturers: the detection of product description texts, and the detection of product names within descriptions from raw text only.

While the results for the text classification and product name detection based on character-level features are promising, the detection of product names from sentences requires further work, especially in creating a relevant training set for this niche problem domain.
This paper mostly focuses on English text, the classification components will have to work in other languages as well, which makes the retrieval of enough training data even more challenging.

\begin{table}[t]
\centering

\caption{
LSTM hyper parameters for POS-based and Character based product name detection. Both LSTM-based classifiers have been prototyped using the pytorch deep learning framework.
}
\label{tab:hyper}

\begin{tabularx}{\columnwidth}{Xrr}

\toprule
Parameter                         &   LSTM-CHAR &    LSTM-POS      \\
\midrule
Hidden Units per layer            &     50      &    70               \\
Number of Layers (stacked LSTM)   &     1       &    3              \\
Dropout                           &     0.1     &    0.1                 \\
Learning Rate                     &     0.1     &    0.01               \\
Epochs                            &     50      &     50              \\
Loss function                    & \multicolumn{2}{r}{Negative log likelihood}   \\
Loss rescaling                    &     1/1     &     1/25             \\

\bottomrule                                
\end{tabularx}                   
\end{table}

\section*{Acknowledgment}

This work is supported by the German Federal Ministry of Education and Research (BMBF), project \emph{CAD-SE}, under grant no. \mbox{01\,IS\,16002\,A/B}.

\vfill

\printbibliography

\end{document}